\documentclass{elsart}
\usepackage{graphicx}
\usepackage{cite}

\def\ifm#1{\relax\ifmmode#1\else$#1$\fi}
\def\x{\ifm{\times}}
  
\def\up#1{$^{#1}$}  \def\dn#1{$_{#1}$}
\def\DAF{DA\char8NE}  
\def\f{\ifm{\phi}}    
\def\ab{\ifm{\sim}}  \def\x{\ifm{\times}}
\def\gam{\ifm{\gamma}}  \def\pic{\ifm{\pi^+\pi^-}}
\def\pt#1;#2;{\ifm{#1\x10^{#2}}}   
\def\to{\ifm{\rightarrow}}    \def\plm{\ifm{\pm}}
\def\ks{\ifm{K_S}} \def\kl{\ifm{K_L}} 
\def\po{\ifm{\pi^0}}  \def\pio{\ifm{\pi^0\pi^0}} 
   \def\plm{\ifm{\pm}}   
\def\kpkm{\ifm{K^+K^-}}  \def\L{\ifm{{\mathcal L}}}
\def\C{\ifm{C}}  \def\P{\ifm{P}}  \def\T{\ifm{T}}

\def\ctwoppmas{\ifm{\frac{\Delta m_1^2}{\sigma_{m}^2}+\frac{\Delta m_2^2}{\sigma_{m}^2}
+\frac{(E_{K_S}-\sum_{i}E_{\gamma_i})^2}{\sigma_E^2}+\frac{(P_x^{K_S}-\sum_iP_x^
     {\gamma_i})^2}{\sigma_{P_x}^2}}}
\def\ctwoppimp{\ifm{\frac{(P_y^{K_S}-\sum_iP_y^
     {\gamma_i})^2}{\sigma_{P_y}^2}+\frac{(P_z^{K_S}-\sum_iP_z^
     {\gamma_i})^2}{\sigma_{P_z}^2}+\frac{(\pi-\vartheta_{\pi\pi}^{*})^2}{\sigma_{\vartheta_{\pi\pi}^{*}}^2}}}

\def\xc{10\up2\x}  \def\xm{10\up3\x}  \def\xcc{10\up4\x}

\makeatletter
\newdimen\z@ \z@=0pt 
\newskip\z@skip \z@skip=0pt plus0pt minus0pt
\def\m@th{\mathsurround=\z@}
\def\ialign{\everycr{}\tabskip\z@skip\halign} 
\def\eqalign#1{\null\,\vcenter{\openup\jot\m@th              
  \ialign{\strut\hfil$\displaystyle{##}$&$\displaystyle{{}##}$\hfil
      \crcr#1\crcr}}\,}

\def\section{\@startsection {section}{1}{\z@}%
                            {-1.5ex plus -1ex minus -.2ex}%
                            {1.3ex plus .2ex}{\normalsize\bf}}
\def\subsection{\@startsection{subsection}{2}{\z@}%
                              {-.8ex plus -.6ex minus -.2ex}
                              {.5ex plus .2ex}{\normalsize}}
\makeatother

\newcommand{\aff}[2]
  {Dipartimento di Fisica dell'Universit\`a #1 e Sezione INFN, #2, Italy.}
\newcommand{\affd}[1]
  {Dipartimento di Fisica dell'Universit\`a e Sezione INFN, #1, Italy.}
\newcommand{\br}{\mbox{${\rm BR}$}}

\newcommand{\AmS}{{\protect\the\textfont2
  A\kern-.1667em\lower.5ex\hbox{M}\kern-.125emS}}

\begin{document}

\begin{frontmatter}

\title{\boldmath A direct search for the \C\P-violating decay $K_S\,$~\to~$\,3\pi^0$
with the KLOE detector at \DAF}

%
\collab{The KLOE Collaboration}
\author[Na]{F.~Ambrosino},
\author[Frascati]{A.~Antonelli},
\author[Frascati]{M.~Antonelli},
\author[Roma3]{C.~Bacci},
\author[Frascati]{P.~Beltrame},
\author[Frascati]{G.~Bencivenni},
\author[Frascati]{S.~Bertolucci},
\author[Roma1]{C.~Bini},
\author[Frascati]{C.~Bloise},
\author[Roma1]{V.~Bocci},
\author[Frascati]{F.~Bossi},
\author[Frascati,Virginia]{D.~Bowring},
\author[Roma3]{P.~Branchini},
\author[Roma1]{R.~Caloi},
\author[Frascati]{P.~Campana},
\author[Frascati]{G.~Capon},
\author[Na]{T.~Capussella},
\author[Roma3]{F.~Ceradini},
\author[Frascati]{S.~Chi},
\author[Na]{G.~Chiefari},
\author[Frascati]{P.~Ciambrone},
\author[Virginia]{S.~Conetti},
\author[Frascati]{E.~De~Lucia},
\author[Roma1]{A.~De~Santis},
\author[Frascati]{P.~De~Simone},
\author[Roma1]{G.~De~Zorzi},
\author[Frascati]{S.~Dell'Agnello},
\author[Karlsruhe]{A.~Denig},
\author[Roma1]{A.~Di~Domenico},
\author[Na]{C.~Di~Donato},
\author[Pisa]{S.~Di~Falco},
\author[Roma3]{B.~Di~Micco},
\author[Na]{A.~Doria},
\author[Frascati]{M.~Dreucci},
\author[Frascati]{G.~Felici},
\author[Karlsruhe]{A.~Ferrari},
\author[Frascati]{M.~L.~Ferrer},
\author[Frascati]{G.~Finocchiaro},
\author[Frascati]{C.~Forti},
\author[Roma1]{P.~Franzini},
\author[Roma1]{C.~Gatti},
\author[Roma1]{P.~Gauzzi},
\author[Frascati]{S.~Giovannella},
\author[Lecce]{E.~Gorini},
\author[Roma3]{E.~Graziani},
\author[Pisa]{M.~Incagli},
\author[Karlsruhe]{W.~Kluge},
\author[Moscow]{V.~Kulikov},
\author[Roma1]{F.~Lacava},
\author[Frascati]{G.~Lanfranchi},
\author[Frascati,StonyBrook]{J.~Lee-Franzini},
\author[Karlsruhe]{D.~Leone},
\author[Frascati]{M.~Martini\corauthref{cor1}},
\author[Na]{P.~Massarotti},
\author[Frascati]{W.~Mei},
\author[Na]{S.~Meola},
\author[Frascati]{S.~Miscetti\corauthref{cor2}},
\author[Frascati]{M.~Moulson},
\author[Karlsruhe]{S.~M\"uller},
\author[Frascati]{F.~Murtas},
\author[Na]{M.~Napolitano},
\author[Roma3]{F.~Nguyen},
\author[Frascati]{M.~Palutan},
\author[Roma1]{E.~Pasqualucci},
\author[Roma3]{A.~Passeri},
\author[Frascati,Energ]{V.~Patera},
\author[Na]{F.~Perfetto},
\author[Roma1]{L.~Pontecorvo},
\author[Lecce]{M.~Primavera},
\author[Frascati]{P.~Santangelo},
\author[Roma2]{E.~Santovetti},
\author[Na]{G.~Saracino},
\author[Frascati]{B.~Sciascia},
\author[Frascati,Energ]{A.~Sciubba},
\author[Pisa]{F.~Scuri},
\author[Frascati]{I.~Sfiligoi},
\author[Frascati]{T.~Spadaro},
\author[Roma1]{M.~Testa},
\author[Roma3]{L.~Tortora},
\author[Roma1]{P.~Valente},
\author[Karlsruhe]{B.~Valeriani},
\author[Frascati]{G.~Venanzoni},
\author[Roma1]{S.~Veneziano},
\author[Lecce]{A.~Ventura},
\author[Roma3]{R.~Versaci},
\author[Frascati,Beijing]{G.~Xu}
\address[Beijing]{Permanent address: Institute of High Energy 
Physics of Academica Sinica, Beijing, China.}
\address[Frascati]{Laboratori Nazionali di Frascati dell'INFN, 
Frascati, Italy.}
\address[Karlsruhe]{Institut f\"ur Experimentelle Kernphysik, 
Universit\"at Karlsruhe, Germany.}
\address[Lecce]{\affd{Lecce}}
\address[Moscow]{Permanent address: Institute for Theoretical 
and Experimental Physics, Moscow, Russia.}
\address[Na]{Dipartimento di Scienze Fisiche dell'Universit\`a 
``Federico II'' e Sezione INFN,
Napoli, Italy}
\address[Pisa]{\affd{Pisa}}
\address[Energ]{Dipartimento di Energetica dell'Universit\`a 
``La Sapienza'', Roma, Italy.}
\address[Roma1]{\aff{``La Sapienza''}{Roma}}
\address[Roma2]{\aff{``Tor Vergata''}{Roma}}
\address[Roma3]{\aff{``Roma Tre''}{Roma}}
\address[StonyBrook]{Physics Department, State University of New 
York at Stony Brook, USA.}
\address[Virginia]{Physics Department, University of Virginia, USA.}
\corauth[cor1]{cor1}{\small $^1$ Corresponding author: Matteo Martini
INFN - LNF, Casella postale 13, 00044    Frascati (Roma), 
Italy; tel. +39-06-94032896, e-mail matteo.martini@lnf.infn.it}

\corauth[cor2]{cor2}{\small $^2$ Corresponding author: Stefano Miscetti
INFN - LNF, Casella postale 13, 00044    Frascati (Roma), 
Italy; tel. +39-06-94032771, e-mail stefano.miscetti@lnf.infn.it}
%
\begin{abstract}
We have searched for the decay $K_S \to 3 \pi^{0}$ with the 
KLOE experiment at DA\char8NE using data from $e^+ e^-$ 
collisions at a center of mass energy
$W\sim m_{\phi}$ for an integrated luminosity \L\ = 450 pb$^{-1}$. 
The search has been performed with a pure \ks\ beam obtained by 
tagging with \kl\ interactions in the calorimeter and 
detecting six photons. We find an upper limit for the branching 
ratio of $1.2 \times 10^{-7}$ at 90\% C.L.
\end{abstract}
\end{frontmatter}
\parindent 6mm

\section{Introduction}
The decay $\ks\to 3\pi^0$ violates \C\P\ invariance. 
The parameter $\eta_{000}$, defined as the  ratio of 
\ks\ to \kl\ decay amplitudes, can be written as: 
$\eta_{000} = A(\ks\to3\po)/A(\kl\to3\po)=
\epsilon + \epsilon'_{000}$, where $\epsilon$ quantifies 
the \ks\ \C\P\ impurity and $\epsilon'_{000}$ is 
due to a direct \C\P-violating term. 
Since we expect $\epsilon'_{000}\;\ll\;\epsilon$ 
\cite{dambrosio}, it follows that $\eta_{000}\!\ab\!\epsilon$.  
In the Standard Model, therefore, 
BR($\ks\to 3\pi^0$) \ab \pt1.9;-9; to an accuracy of a few \%, 
making the direct observation 
of this decay quite a challenge. 

The best upper limit on BR($\ks\to 3\pi^0$) from a search 
for the decay was obtained by the SND experiment at 
Novosibirsk. They find 
BR($\ks\to 3\pi^0$)\pt\,\leq1.4;-5; at 90\% C.L. 
\cite{SND3pi0}.
CPLEAR has pioneered the method of searching for 
interference between \ks\ and \kl\ decays. Interference 
results in the appearance of a 
term $\Re\,(\eta_{000})\cos(\Delta m\,t)-\Im\,(\eta_{000})\sin(\Delta m\,t)$
in the decay intensity. $\Re\,(\eta_{000})$ 
and $\Im\,(\eta_{000})$ are obtained from a fit, 
without discriminating between \kl\ or $\ks\to\ 3\po$ decays. In this 
way CPLEAR finds $\eta_{000}=(0.18\pm0.15)+i\,(0.15\pm 0.20)$
\cite{CPLEAR3pi0}.
The NA48 collaboration~\cite{NA48} has recently 
reached much higher sensitivity. By fitting the 
$\ks/\kl\to 3\pi^0$ interference pattern at small decay 
times, they find 
$\Re\,(\eta_{000})=-0.002\;\pm\;0.011_{\rm stat}\;\pm\;0.015_{\rm sys}$ 
and $\Im\,(\eta_{000})=-0.003\;\pm\;0.013_{\rm stat}\;\pm\;0.017_{\rm sys}$, 
corresponding to BR($\ks\to 3\pi^0$)\pt\,\leq7.4;-7; at 90\% C.L. 
The sensitivity to \C\P\T\ violation via unitarity \cite{Zhou} is now 
limited by the error in 
$\eta_{+-} = A(\kl \to \pi^+\pi^-)/A(\ks \to \pi^+\pi^-)$. 

We report in the following an improved limit from a 
{\it direct search} for the 3\po\ decays of the \ks.
Apart from the interest in confirming the Standard Model, knowledge 
of $\eta_{000}$ allows tests of the validity of 
\C\P\T\ invariance using unitarity.

\section{\DAF\ and KLOE}
The data were collected with the KLOE detector~[\citen{kloe1}--\citen{kloe4}]
at DA$\Phi$NE~\cite{dafne},
the Frascati $\phi$ factory. DA$\Phi$NE is an 
$e^+e^-$ collider
operated at a center-of-mass energy $W\sim1020$ MeV, 
the mass of the $\phi$ meson. Positron and electron 
beams of equal energy collide at an angle of $\pi-$ 
0.025 rad, producing $\phi$ mesons nearly at rest 
($p_\phi$ \ab\ 12.5 MeV). $\phi$ mesons decay 34\% of the 
time into nearly collinear $K^{0}\overline K^0$ pairs.
Because $J^{PC}(\phi)=1^{--}$, the kaon pair is in a 
\C-odd antisymmetric state, so that the final state is 
always \ks-\kl. Detection of a \kl\ signals the 
presence of a \ks\ of known momentum and direction. 
We say that detection of a \kl\ ``tags'' the \ks.

The KLOE detector consists of a large cylindrical 
drift chamber (DC), surrounded by a 
lead/scintillating-fiber electromagnetic calorimeter 
(EMC). A superconducting coil around the calorimeter 
provides a 0.52 T field. 
The drift chamber, 4~m in diameter and 3.3~m long, 
is described in Ref. \citen{kloe1}. The momentum 
resolution is $\sigma(p_{\perp})/p_{\perp}\approx 0.4\%$. 
Two track vertices are reconstructed with a spatial 
resolution of \ab\ 3~mm. 
The calorimeter, described in Ref. \citen{kloe2}, 
is divided into a barrel and two endcaps, for a 
total of 88 modules, and covers 98\% of the solid 
angle. The modules are read out at both ends by 
photomultipliers providing energy deposit and arrival 
time information. 
The readout segmentation provides 
the coordinates transverse to the fiber plane.
The coordinate along the fibers is 
obtained by the difference between the arrival times of the
signals at either end. 
Cells close in time and space are grouped into 
calorimeter clusters. The energy and time resolutions 
are $\sigma_E/E = 5.7\%/\sqrt{E\ {\rm(GeV)}}$ and  
$\sigma_T = 54\ {\rm ps}/\sqrt{E\ {\rm(GeV)}}\oplus50\ {\rm ps}$, 
respectively.

The KLOE trigger, described in Ref. \citen{kloe4}, 
uses calorimeter and chamber information. For this 
analysis, only the calorimeter signals are used. 
Two energy deposits above threshold ($E>50$ MeV for the 
barrel and  $E>150$ MeV for the endcaps) are required. 
Recognition and rejection of cosmic-ray events is also 
performed at the trigger level. Events with two energy 
deposits above a 30 MeV threshold in two of the outermost 
calorimeter planes are rejected.

During 2002 data taking, the maximum luminosity 
reached by DA$\Phi$NE was \pt7.5;31; cm\up{-2}s\up{-1}, 
and in September 2002, \DAF\ delivered 91.5 pb\up{-1}. 
We collected data in 2001-2002 for an integrated 
luminosity \L\ = 450 pb$^{-1}$. A total of 1.4 billion $\phi$ mesons 
were produced, yielding 450 million \ks-\kl\ pairs. 
Assuming BR($\ks\to 3\pi^0$) = \pt1.9;-9;, \ab1 signal 
event is expected to have been produced.

The mean decay lengths of the $K_S$ and $K_L$ are 
$\lambda_S \sim 0.6$ cm and $\lambda_L \sim 340$ cm 
at DA$\Phi$NE. 
About 50\% of $K_L$'s reach the calorimeter before 
decaying. The $K_L$ interaction in the calorimeter 
(``\kl\ crash'') is identified by requiring a cluster 
with energy greater than 100 MeV that is not associated 
to any track and whose time corresponds to a velocity 
in the $\phi$ rest frame, $\beta^*$, of 
\ab\ 0.2. The \kl-crash provides a very 
clean $K_S$ tag. The average value of the center-of-mass 
energy, $W$, is obtained with a precision of 30 keV  
for each 100 nb$^{-1}$  running period (of duration $\sim 1$ hour)
using large-angle Bhabha events. The value of $W$ and 
the $K_L$-crash cluster position allows us to establish, 
for each event, the trajectory of the $K_S$ with an 
angular resolution of 1$^{\circ}$ and a momentum 
resolution better than 2 MeV.

Because of its very short lifetime, the displacement
of the $K_S$ from the $\phi$ decay position 
is negligible. We therefore identify as \ks\ decay 
photons neutral particles that travel with $\beta=1$ 
from the interaction point (IP) to the EMC. Each cluster 
is required to satisfy the condition 
$|T-R/c|<{\rm min}(3\sigma_T, 2\ {\rm ns})$, 
where $T$ is the photon flight time and $R$ the path 
length; $\sigma_T$ also includes a contribution from 
the finite bunch length (2--3 cm), which introduces 
a dispersion in the collision time. 

In order to retain a large control sample for the 
background while preserving high efficiency for the 
signal, we keep all photons satisfying $E_\gamma>$ 7 ~MeV and
$|\cos(\theta)|<$ 0.915. The photon detection efficiency 
is \ab90\% for $E_\gamma$ = 20 MeV, and reaches 
100\% above 70 MeV. The signal is searched for by requiring six
prompt photons after tagging.

The normalization is provided by counting the $\ks\to 2\pi^0$ events
in the same tagged sample.

\section{Monte Carlo simulation}
The response of the detector to the decay of interest and 
the various backgrounds is studied using the 
Monte Carlo (MC) program GEANFI \cite{NIM}.  
GEAN\-FI accounts for 
changes in machine operation and background conditions,
following the machine conditions run by 
run, and has been calibrated with Bhabha scattering events and  other 
processes. The response of the EMC to \kl\ interactions 
is not simulated but has been obtained from a large 
sample of \kl-mesons tagged by identifying 
$\ks\to\pic$ decays. 
This not only gives accurate representation of the 
EMC response to the 
$K_L$ crash, but also results in an effective 40\%
increase in MC statistics.
The \kl-crash 
efficiency cancels in the final 3\po/2\po\ ratio to 
better than 1\% and we assign a 0.9\% systematic error 
to the final result due to this source.

Backgrounds are obtained from MC $\f \to \ks\kl$ events 
corresponding to an integrated luminosity \L\ = 900 pb\up{-1}. 
We also use a MC sample of $\f\to\kpkm$ 
events for \L\ = 450 pb$^{-1}$ and a MC sample of radiative \f\ 
decays for \L\ = 2250 pb$^{-1}$. A sample of \ab\ $340\,000$ 
$\ks\to 3\pi^0$ 
MC events is used to obtain the signal efficiency.
\section{Photon counting for data and Monte Carlo}
To test how well the MC reproduces the observed photon 
multiplicity after tagging, we determine the fraction 
of events of given  multiplicity, $N_{\gamma}=k$, 
defined as 
$F(k)= N_{{\rm ev}}(N_{\gamma}=k)/\sum_{i=3}^{6}{N_{{\rm ev}}(N_{\gamma}=i)}.$
As shown in Table \ref{tab:fraction}, there is a 
significant discrepancy between data and Monte Carlo 
for events with multiplicity five and six.
\begin{table*}[hb]
\begin{center}
\newcommand{\m}{\hphantom{$-$}}
\newcommand{\cc}[1]{\multicolumn{1}{c}{#1}}
\renewcommand{\arraystretch}{1.1} 
\begin{tabular}{@{}lcccc}
\hline
     &    Data 2001 &  MC 2001 &  Data 2002 &  MC 2002 \\ \hline 
$F$(3) &30.95\plm0.16 &30.31\plm0.11 &30.79\plm0.12 &30.06\plm0.08 \\
$F$(4) &67.35\plm0.23 &67.93\plm0.17 &67.93\plm0.18 &68.15\plm0.12 \\
$F$(5) & 1.55\plm0.01 & 1.80\plm0.01 & 1.19\plm0.01 & 1.66\plm0.01 \\
$F$(6) & 0.15\plm0.01 & 0.14\plm0.01 & 0.08\plm0.01 & 0.13\plm0.01 \\ \hline
\hline
\end{tabular}\\[2pt]
\caption{Measured values of $F$ for data and Monte Carlo samples, in percent. }
\label{tab:fraction}
\end{center}
\end{table*}
These samples
are dominated by $K_S \to 2 \pi^0$ decays plus 
additional clusters  due either  to shower fragmentation 
({\em split clusters}) or the accidental coincidence 
of machine background photons ({\em accidental clusters}).
To understand this discrepancy, 
we have measured the probability, $P_A(1,2)$, 
of having one, or more than one,
accidental cluster passing our selection
by extrapolating
the rates measured in an out-of-time window, 
$(-68 \leq T \leq -14)$~ns, 
that is earlier than the bunch crossing. 
In Table \ref{tab:probacci_split}, 
we list the average values of these probabilities.
\begin{table*}[hb]
\begin{center}
\newcommand{\m}{\hphantom{$-$}}
\newcommand{\cc}[1]{\multicolumn{1}{c}{#1}}
\renewcommand{\arraystretch}{1.2} 
\begin{tabular}{@{}lcccc}
\hline
$F(K)$         &   Data 2001 &  MC 2001    &  Data 2002  &  MC 2002 \\ \hline
\xc$P_A$(1) &0.75\plm0.30 &1.03\plm0.16 &0.38\plm0.17 &0.89\plm0.08 \\
\xc$P_A$(2) &0.14\plm0.05 &0.16\plm0.03 &0.07\plm0.02 &0.10\plm0.03 \\ 
\xm$P_S(1)$    &3.6 \plm 0.2 &3.8 \plm 0.3 &3.7 \plm 0.2 & 3.3 \plm 0.1 \\
\xcc$P_S(2)$   &1.5 \plm 0.4 &1.5 \plm 0.3 &0.9 \plm 0.2 & 1.7 \plm 0.2 \\ \hline
\hline
\end{tabular}\\[2pt]
\caption{Measured values of the probabilities 
$P_A$ and $P_S$.}
\label{tab:probacci_split}
\end{center}
\end{table*}
The observed discrepancy has been traced to an 
understood problem with the procedure for the selection 
of machine-background clusters. 

The MC-true fraction of 
events with a given multiplicity, $f_{MC}$, is obtained 
by ignoring clusters due to machine background and 
counting at most one cluster per simulated particle 
incident on the calorimeter.
Using the fractions $f_{MC}$, together with the 
values of $P_A$ obtained as discussed above, 
we fit the observed $F(k)$ distribution to get the
probability for a cluster to generate fragments, $P_{S}$ 
(see Table \ref{tab:probacci_split}).
This fit accurately reproduces the observed
fractions in the multiplicity bins five and six. 
More details on these measurements
can be found in Ref. \citen{NotaAcci}. 
The results of this study demonstrate the need for 
careful calibration of the background composition 
when comparing data and MC samples.

\section{ Data analysis}
$\ks\to 3\pi^0$ candidates consist of a \kl\ crash plus six photons. 
In our data sample of \L\ = 450 pb\up{-1}, we find $39\,538$ events, 
essentially all background. After removing background, we 
obtain the branching ratio by normalizing to the number of 
$\ks\to 2\pi^0$ events. The latter are found by asking for three 
to five prompt photons plus the \kl-crash.

According to the MC, the six-photon sample is dominated (95\%) 
by $\ks\to 2\pi^0$ decays plus two additional photon clusters. 
These clusters are due to fragmented or split showers 
(2S, 1S+1A, 34\%) and to accidental photons from machine background 
(2A, 64\%). 
About 2\% of the background events are due to false $K_L$-crash
tags from $\f\to\ks\kl\to\pic$, 3\po\ events.
In such events, charged pions from \ks\ decays interact in the low-beta 
insertion quadrupoles\footnote{The first quadrupoles are located 
approximately 45 cm on either side of the IP.}, ultimately 
simulating the \kl-crash signal, while \kl\ decays close to 
the IP produce six photons.
Similarly, $\f\to\kpkm$ events give a false signal (\ab\ 1\%), 
as well as $\f\to\eta\gam\to 3\pi^0\gam$ events (\ab\ 0.3\%).
The cuts described in the following make these contaminations 
negligible.

\begin{figure}[htb]
\begin{center}
\includegraphics[width=0.5\textwidth]{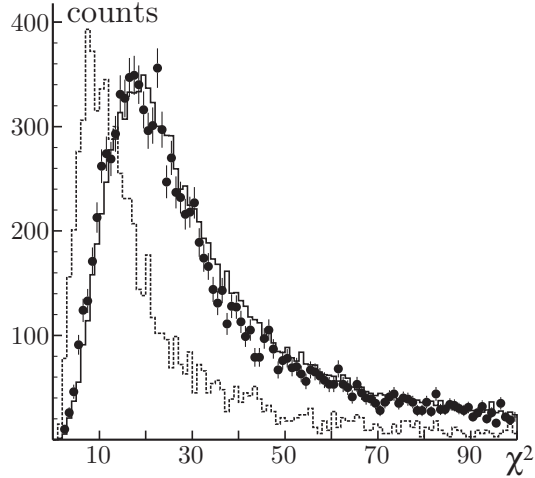}
\vglue-15mm\end{center}
\caption{Distribution of $\chi^2$ for 
the tagged six-photon sample for data (points),
MC background (solid line), and $10\,000$ events of MC signal (dashed line).}
\label{fig:chi2fit}
\end{figure}

\begin{figure}[hb]
\begin{center}
\includegraphics[width=1.\textwidth]{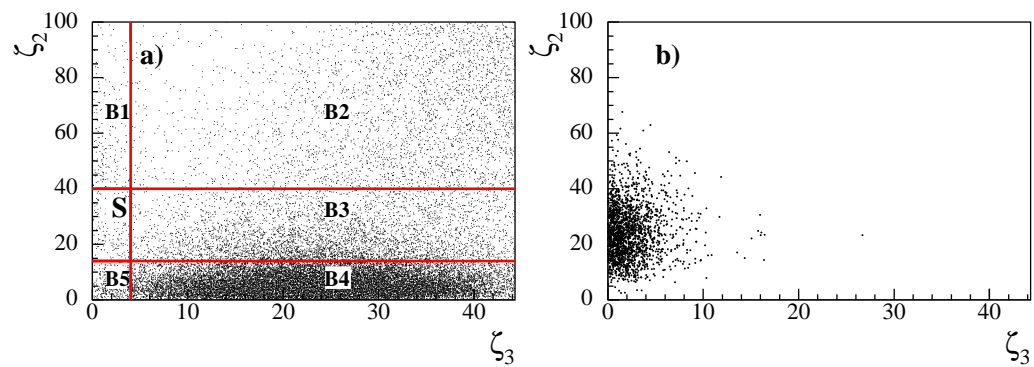}
\end{center}
\vspace{-.3cm}
\caption{Scatter plot of $\zeta_2$ vs. 
$\zeta_3$ plane for the tagged six-photon sample:
a) MC background, b) MC signal.}
\label{fig:scatterplots}
\end{figure}

To reduce the background, we first perform a kinematic fit 
with 11 constraints: energy and momentum conservation, the 
kaon mass and the velocity of the six photons. 
The $\chi^2$ distribution of the fit to data and MC background 
is shown in Fig. \ref{fig:chi2fit}.
In the same plot, we also show the expected shape for 
signal events.
Cutting at a reasonable $\chi^2$ value 
($\chi^{2}/\rm{11} < 3$)
retains 71\% of the signal while considerably
reducing the background from false $K_L$-crash events
(33\%), in which the direction of the $K_S$
and $K_L$ are not correlated.
However, this cut is not as effective on
the 2S, 2A background, due to
the soft energy spectrum of the fake  clusters.
In order to gain rejection power over the background for
events with split and accidental clusters,
we look at the correlation between the following two 
$\chi^2$-like estimators:
\begin{itemize}
\item $\zeta_2$, defined as 
\begin{eqnarray}
\zeta_2 & = & \ctwoppmas+ \nonumber \\
              && \;\;\; + \ctwoppimp \nonumber \\
\nonumber
\end{eqnarray}
selecting the four out of six photons 
that provide the
best kinematic agreement with the $K_S\to 2\pi^0$ decay 
hypothesis. This variable is quite insensitive to fake 
clusters. It is constructed using the two values of 
$\Delta m$ = $m_i-m_{\pi^0}$ (where $m_i$ is the 
invariant mass of a photon pair), 
the opening angle between $\pi^0$'s in the $K_S$ rest 
frame, and 4-momentum conservation.
The resolutions on these quantities have been 
evaluated using a control sample of events with a 
$K_L$-crash and four prompt photons.
\item $\zeta_3$, defined as 
$$\zeta_3=\frac{\Delta m_1^2}{\sigma_{m}^2}+
\frac{\Delta m_2^2}{\sigma_{m}^2}+
\frac{\Delta m_3^2}{\sigma_{m}^2}$$
where the pairing of the six photons into $\pi^0$'s is
performed by minimizing
this variable. 
$\zeta_3$ is close to zero for a $\ks\to 3\pi^0$ event and 
large for six-photon background events. 
\end{itemize}
For each estimator, the photon pairing 
with smallest $\zeta$ value is kept. 
Figure~\ref{fig:scatterplots}a shows the distribution of 
events in the $\zeta_2$-$\zeta_3$ plane 
for the MC background. Most of the events are 
concentrated at low values of $\zeta_2$,
as expected for $\ks\to 2\pi^0$ events plus some additional 
isolated energy deposits in the EMC.
A clear signal/background separation is achieved as 
can be seen by comparing the background and signal
distributions in Figs.~\ref{fig:scatterplots}a and 
\ref{fig:scatterplots}b.
\begin{figure}[htb]
\begin{center}
\begin{tabular}{cc}
\includegraphics[width=.8\textwidth]{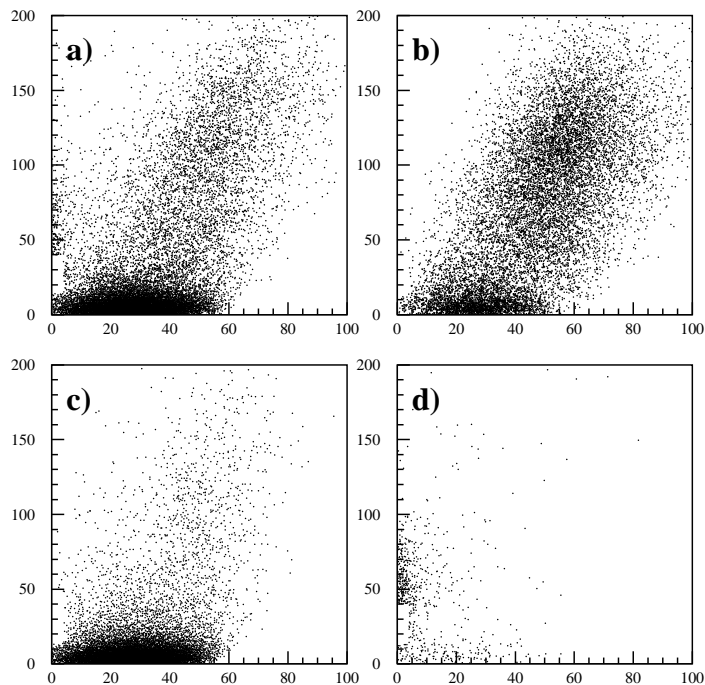}
\end{tabular}
\end{center}
\caption{Scatter plots of $\zeta_2$ vs. $\zeta_3$ for
the tagged six-photon sample:
data (a), MC sample with two split clusters (b),
two accidental clusters (c), and false $K_L$-crash events (d).}
\label{fig:chi2chi3fit}
\end{figure}
We subdivide the $\zeta_2$-$\zeta_3$ plane into the six regions 
B1, B2, B3, B4, B5, and S as indicated in Fig. 
\ref{fig:scatterplots}a. Region S, with the largest 
signal-to-background value, is the ``signal'' box.

\begin{table*}[htb]
\newcommand{\m}{\hphantom{$-$}}
\renewcommand{\arraystretch}{1.2} 
\begin{tabular}{@{}lcccccc}
\hline
     &  B1  &  B2 & {\bf S} & B3 &  B5 &  B4 \\ \hline
data    & $452\pm 21$ & $10132\pm 101$ & 
          {\boldmath $282\pm 17$} & $5037\pm 71$   & 
          $326\pm 18$ & $22309\pm 149$ \\
MC      & $419\pm 19$ & $9978\pm 104$ & 
          {\boldmath $282\pm 13$} & $4816\pm 43$ & 
          $380\pm 10$ & $22682\pm 190$\\
\hline
\end{tabular}\\[2pt]
\caption{Comparison between data and MC expectations
in the different regions of the $\zeta_2$-$\zeta_3$ plane 
for the entire sample
with a $K_L$-crash and six prompt photons. The boxes are defined as in 
Fig.~\ref{fig:scatterplots}a.}
\label{tab:compare}
\end{table*}
The scatter plot in the $\zeta_2$-$\zeta_3$ plane for 
the data is shown in Fig.~\ref{fig:chi2chi3fit}a.
Our MC simulation does not accurately reproduce the 
absolute number of 2S and 2A background events. This 
is also true of the predicted number of false 
$K_L$-crash events. However, the simulation does 
describe the kinematical properties of these events 
quite well.
The two-dimensional $\zeta_2$-$\zeta_3$ distribution 
allows us to calibrate the contributions from the different backgrounds. 
The MC shapes for each of the three categories are 
shown in Figs.~\ref{fig:chi2chi3fit}b-d.
We perform a binned likelihood fit of a linear combination 
of these shapes to the data, excluding the signal-box 
region. From the fit we find the composition of 
the six-photon sample to be (37.9\plm1.0)\%, (57.4\plm1.3)\%,  
and (4.7\plm0.3)\% for the 2S, 2A, and false $K_L$-crash 
categories, respectively.

\begin{figure}[!t]
\begin{center}
\includegraphics[width=.6\textwidth]{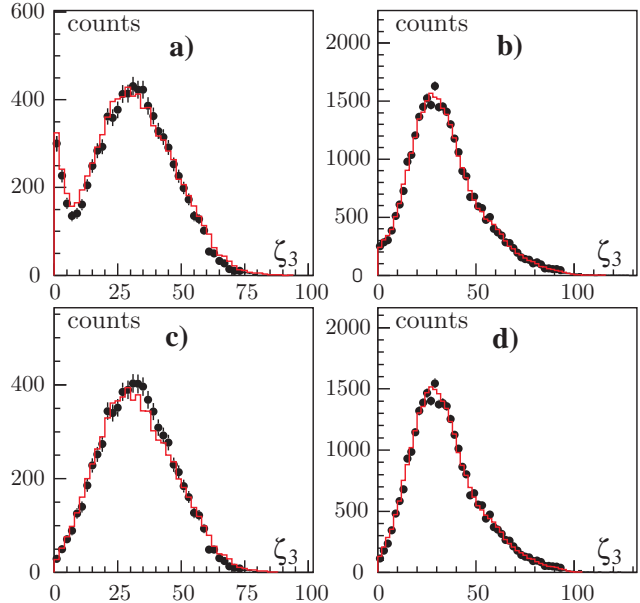}
\end{center}
\caption{
Distributions in $\zeta_3$ for the tagged six-photon sample.
Plots on the left are for events in the central band in
$\zeta_2$; 
plots on the right are for events in all other regions of 
the plane. 
The plots in the top row are for the entire sample, 
before any cuts are made. The plots in the bottom row are after
the application of the track veto.
In all cases, black points represent data; solid line represents MC.}
\label{fig:chi2chi3}
\end{figure}

As a check, we compare data and MC for the projected 
distribution in $\zeta_3$ for the three bands in $\zeta_2$,
as shown in 
Figs.~\ref{fig:chi2chi3}a-b. Excellent  agreement is observed. 
The large peak at low values of $\zeta_3$ in the central band 
is due to the false $K_L$-crash events.
As a final test, we compare data and MC in the signal box 
and the five surrounding control regions. The agreement is 
better than 10\% in all regions, as seen from Table \ref{tab:compare}.

Although cutting on $\chi^2$ substantially suppresses 
the false $K_L$-crash background, we reduce this background 
to a negligible level by vetoing events with tracks coming 
from the IP. This effectively eliminates events in which 
the false $K_L$-crash is due to a $\ks\to \pic$ decay with 
the pion secondaries interacting in the quadrupoles. The 
effect on the signal region can be appreciated by comparison 
of Figs.~\ref{fig:chi2chi3}a and \ref{fig:chi2chi3}c. 
Moreover, in order to improve the quality of the photon 
selection using $\zeta_2$, we cut on the variable 
$\Delta E=(m_\phi/2-\sum E_i)/\sigma_{E}$, where $i$ = 1--4 stands 
for the four chosen photons 
in the $\zeta_2$ estimator
and $\sigma_E$ is the appropriate 
resolution. For $\ks\to 2\pi^0$ decays plus two background 
clusters, we expect $\Delta E$ \ab\ 0, 
while for $\ks\to\pio\po$, $\Delta E\geq m_{\pi^0}/\sigma_{E}$. 

Before opening the signal box, 
we refine our cuts on $\chi^2$, $\zeta_2$, $\zeta_3$, 
and  $\Delta E$ 
using the optimization procedure described in Ref.~\citen{METODO}.
\begin{figure}[!t]
\begin{center}
\includegraphics[width=.8\textwidth]{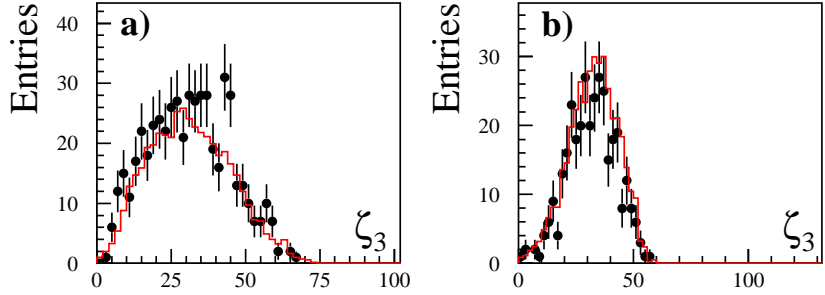}
\end{center}
\caption{Distributions of $\zeta_3$ for 
 the central band 12.1$<\zeta_2<$60 (a),
 the side-bands $\zeta_2<$12.1, $\zeta_2> $60 (b),
 after all cuts.
 Points represent data, solid line MC.}
\label{fig:results}
\end{figure}
We end up choosing $\chi^2<40.4$ and $\Delta E>$1.7. 
The signal box is defined by 12.1 $<\zeta_2<$ 60 and $\zeta_3<$ 4.6. 

Figures \ref{fig:results}a and \ref{fig:results}b show the $\zeta_3$ distributions
for the central band and the sidebands in $\zeta_2$.

\begin{figure}[htb]
\begin{center}
\includegraphics[width=.6\textwidth]{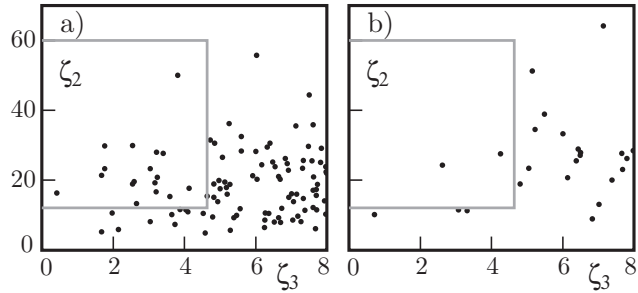} 
\end{center}
\caption{Distribution of $\zeta_2$ vs $\zeta_3$ 
after cuts: MC background 
900 pb$^{-1}$ (a), data 450 pb$^{-1}$ (b).}
\label{fig:scatter_end}
\end{figure}

\begin{table*}[t]
\begin{center}
\newcommand{\m}{\hphantom{$-$}}
\renewcommand{\arraystretch}{1.1} 
\begin{tabular}{@{}lcccccc}
\hline
     & B1 &  B2 &  {\bf S} &  B3 &  B4 &  B5 \\ \hline
data & 0  & 4 \plm\ 2&{\bf 2.0 \plm\ 1.4}&520 \plm\ 23&3 \plm\ 2&326 \plm\ 18\\
MC   &0&3.2 \plm\ 0.8&{\bf3.1 \plm\ 0.8}&447 \plm\ 10&2.5 \plm\ 0.8&389 \plm\ 10\\ \hline
\end{tabular}\\[2pt]
\end{center}
\caption{Same as Table \ref{tab:compare}, after all cuts.}
\label{tab:compare_fin}
\end{table*}
In Table~\ref{tab:compare_fin}, we also list 
the number of events obtained in each of the six regions
of the $\zeta_2$-$\zeta_3$ plane 
at this final stage of the analysis.
In Figs.~\ref{fig:scatter_end}a-b we show the 
$\zeta_2$-$\zeta_3$ scatter plots
for data and Monte Carlo.
The rectangular region illustrates 
the boundaries of the optimized signal box.
Seventeen MC events are counted 
in this
region before applying the data-MC scale factors
resulting from the calibration procedure described above.
Contributions to the scale factors include the fact
that the simulated integrated luminosity is greater than
that for the data set (\x2), the increased
$K_L$-crash efficiency in the simulation (\x1.4),
and the increased probability of having accidental or
split clusters in the simulation (on average,
\ab\x1.9).

The selection efficiency at each step of the analysis has been studied 
using the MC.
After tagging, the efficiency for the six-photon selection is
$(47.8 \pm 0.1_{\rm{stat}})$\%.  
Including all cuts, we estimate a total efficiency
of  $\epsilon_{3\pi} = (24.4 \pm 0.1_{\rm{stat}} $)\%. At the
end of the analysis chain,  we have two candidates 
with an expected background of 
$B^{\rm{exp}} = 3.13 \pm 0.82_{\rm{stat}}$.

In the same tagged sample, we also count 
events with 
photon multiplicities of three, four, or five.
The corresponding efficiency
is (91.8\plm0.2\dn{\rm stat})\% for $\ks\to\pio$ events. 
The residual background contamination is estimated to be
(0.77 $\pm 0.24_{\rm{stat+sys}})\%$ and 
(0.65$\pm 0.10_{\rm{stat+sys}})\%$ 
in the 2001 and 2002 running periods respectively. 
Subtracting
the background and correcting for the efficiency,
we count \pt3.78;7; $\ks\to\pio$  events. We use this number to normalize 
the number of signal events when obtaining the 
branching ratio.

\section{Systematic uncertainties}
Systematics arise from uncertainties in estimation 
of the acceptance,  backgrounds, and the analysis efficiency. 
The evaluation of the systematic uncertainties is described
in detail in Ref. \citen{NOTA}.

Concerning the acceptance of the event selection for both 
the 2\po\ and 3\po\ samples, we estimate the systematic 
errors in photon counting by comparing  data and MC 
values for the $P_A$ and $P_S$ probabilities described above.
The photon 
reconstruction efficiency 
for both data and MC is evaluated using a large sample of 
$\f\to\pic\po$, \po\to\gam\gam events. 
The momentum of one of the photons is estimated from 
tracking information and position of the other cluster.
The candidate photon is searched for within a search
cone.
The efficiency is 
parameterized as a function of the photon energy. 
Systematics related to this correction are obtained
from the variation of the efficiency as a function
of the width of the search cone.
The results are listed in Table \ref{tab:syst_effi} 
under the heading {\em cluster}. 
\begin{table*}[!htb]
\begin{center}
\newcommand{\m}{\hphantom{$-$}}
\newcommand{\cc}[1]{\multicolumn{1}{c}{#1}}
\renewcommand{\arraystretch}{1.1} 
\begin{tabular}{@{}lcc}
\hline
  &    $\Delta \alpha / \alpha$ 
 ($K_S \to 2 \pi^0$) & 
  $\Delta \alpha / \alpha$ ($K_S \to 3 \pi^0$)
 \\ \hline
Cluster  &  0.16 \%   & 0.70 \% \\
Trigger  &  0.08 \%   & 0.08 \% \\
Background filter   &  0.20 \%   & 0.08 \% \\ \hline \hline
Total    &  0.27 \%   & 0.71 \% \\
\hline
\end{tabular}\\[2pt]
\caption{Systematic acceptance uncertainties, 
$\Delta\alpha$,  for the 2\po\ and 3\po\ event selection criteria.} 
\label{tab:syst_effi}
\end{center}
\end{table*}
The total uncertainty  is smaller for the normalization
sample since an inclusive
selection criterion is used in this case.

The normalization sample also suffers 
a small ($0.4\%$) loss due to the 
use of a filter during data reconstruction to reject
cosmic rays, Bhabha fragments from the low-beta quadrupole,
and machine background events. This loss
is estimated using the MC. We correct for it and
add a 0.2\% systematic error
to the selection efficiency. 
The trigger and cosmic-ray veto efficiencies 
have been estimated with data for the normalization
sample and extrapolated by MC to the signal sample. 
These efficiencies are very close to unity and 
the related systematics  are negligible.

For the tagged six-photon sample, we have investigated 
the uncertainties related to the estimate of the 
background in the signal box after all cuts, 
$B^{\rm{exp}}$. 
We have first considered the error related to the 
calibration of the MC background composition 
by propagating the errors on the scale
factors obtained from the fit. 
This corresponds to a relative error of $2.4\%$ on $B^{\rm{exp}}$.
Moreover, we have investigated the extent to which the 
track-veto efficiency influences the residual false $K_L$-crash
contamination. To do so, we examine the data-MC ratio, $R_{\beta}$, 
of the sidebands in $\beta^{*}$ for events rejected by
this veto, since for true $K_L$'s $\beta^*$ peaks at \ab\ 0.2 while 
false $K_L$-crashes are broadly distributed in $\beta^*$.
 We obtain $R_{\beta}=1.10 \pm 0.01$. 
Knowing that in the MC only 24\% of the fakes survive the 
veto, we find a fractional error of 32\% 
on the fake background. Since false $K_L$-crash
events account for 15\% of the total background,
the error on $B^{\rm{exp}}$ from data-MC differences
in the track veto efficiency is 4.6\%.
 
A control sample of  $K_S \to 2 \pi^{0}$ 
with four prompt photons has been used to compare
the energy scale and 
resolution of the calorimeter in data and in the MC.
The distributions of the $m$ and $\Delta E$ variables have
also been compared by fitting them with Gaussians.
By varying the mass and energy resolution by $\pm 1 \sigma$
in the definitions of $\zeta_2$ and $\zeta_3$, we
observe a relative change of 6.6\% in the background
estimate. Similarly, correcting
for small differences in the energy scale for data and MC, we
derive a systematic uncertainty of 6.7\% on $B^{\rm{exp}}$.

Finally, we have  tested the effect of the cut on
$\chi^2$ by constructing the ratio between 
the cumulative distributions for data and MC. An error
of 5\% is obtained. A summary of all the 
systematic errors on the background estimate is given
in Table~\ref{tab:syst_bkg_ana}.
\begin{table*}[htb]
\begin{center}
\newcommand{\m}{\hphantom{$-$}}
\newcommand{\cc}[1]{\multicolumn{1}{c}{#1}}
\renewcommand{\arraystretch}{1.1} 
\begin{tabular}{@{}lcc}
\hline
  &  $\Delta B^{\rm{exp}} / B^{\rm{exp}}$  & 
    $\Delta \epsilon_{\rm ana} / \epsilon_{\rm ana}$  \\ \hline
Background composition       &  2.4\%    &  -        \\
Track veto        &  4.8\%   &  0.2\%   \\
Energy resolution &  6.6\%        & 0.5\%  \\
Energy scale      &  6.7\%  & 1.0\% \\ 
$\chi^2$     &  5.0\%  &     1.8\%  \\ \hline \hline
Total       &  11.5\%   & 2.1\% \\
\hline 
\end{tabular}\\[2pt]
\caption{Systematic uncertainties on the expected background 
and analysis efficiency, $\epsilon_{\rm ana}$.}
\label{tab:syst_bkg_ana}
\end{center}
\end{table*}
Adding in quadrature all sources we obtain a total
systematic error of 12\%
on the background estimate. 

To determine the systematics related
to the analysis cuts for the signal, we have
first evaluated the effect of the track veto.
Using the MC signal sample, we estimate a 
vetoed event fraction of (3.7 $\pm$ 0.1)\%.
The data-MC ratio of the cumulative distributions
for the track-vetoed events in the tagged six-photon
sample is $R_{TV}=1.06\pm0.03$, which translates into a 
0.2\% systematic error on the track-veto efficiency.

Because of the similarity of the $\chi^2$
distributions 
for the tagged four- and six-photon samples, as confirmed 
by MC studies, an estimate of the systematic error
associated with the application of the $\chi^2$
cut can be obtained from the data-MC comparison
of the cumulative $\chi^2$ distributions for the
four-photon sample.
The systematic error arising from data-MC discrepancies
in the $\chi^2$ distribution is 
estimated to be 1.8\% by this comparison.

Moreover, the  efficiency changes related to
differences between the calorimeter resolution and energy
scale for data and MC events have 
been studied in a manner similar to
that previously described for the evaluation of the 
systematics on
the background. The systematic 
uncertainties on the analysis efficiency are
summarized in Table~\ref{tab:syst_bkg_ana}. Adding all sources
in quadrature we quote a total systematic error of 2.1\%
on the estimate of the analysis efficiency.

\section{Results}
At the end of the analysis, we find 2 events in the signal box 
with an estimated background of 
$B^{\rm{exp}}=3.13 \pm 0.82_{\rm{stat}} \pm 0.37_{\rm{syst}}$. 
To derive an upper limit on the number of signal counts, 
we build the background probability distribution function,
taking into account our finite MC statistics and 
the uncertainties on the MC calibration factors. This 
function is folded with a Gaussian  of width equivalent 
to the entire systematic uncertainty on the background.
Using the Neyman construction described in Ref.~\citen{FC}, 
we limit the number of $\ks\to3\po$ decays observed to 3.45 
at 90\% C.L., with a total reconstruction efficiency 
of $(24.36\pm 0.11_{\rm{stat}}\pm 0.57_{\rm{sys}})\%$. 
In the same tagged sample, we count 
\pt3.78;7; $\ks\to\pio$ events. This number
is used for normalization.
Finally, using the value BR($\ks\to\pio$) = 0.3105 \plm\ 0.0014~\cite{pdg_2004} we obtain:
\begin{equation}
\br(\ks\to3\po) \leq 1.2 \times 10^{-7} \; \;\;\; {\rm at}\;\; \; 90\%\;\;\; {\rm C.L.},
\end{equation}
which represents an improvement by a factor of $\sim$6
with respect to the best
previous limit \cite{NA48}, and
by a factor of 100 with respect to the best limit obtained with
a direct search~\cite{SND3pi0}. 

The limit on the BR can be directly translated into
a limit on $|\eta_{000}|$: 
\begin{equation}
\eqalign{
|\eta_{000}| = &\left|{A( K_S\to3\pi^0)\over A( K_L\to3\pi^0)}\right|=\cr
&\kern2cm=\sqrt{{\tau_L\over\tau_S}\:{{\rm BR}(\ks\to3\po)\over{\rm BR}(\kl\to3\po)}} 
<0.018\ {\rm at}\ 90\%\ {\rm C.L.}\cr}
\end{equation}

This result describes a circle of radius 0.018 centered 
at zero in the $\Re({\eta_{000}})$, $\Im{(\eta_{000})}$ plane and 
represents a limit $2.5$ times smaller
than the result of Ref.~\citen{NA48}. 
As follows from the discussion in that reference, our
result confirms that the sensitivity of the \C\P\T\ test 
from unitarity is now limited by the uncertainty on $\eta_{+-}$.

\section*{Acknowledgments}
%
We thank the DA$\Phi$NE team for their efforts in maintaining 
low-background running 
conditions and their collaboration during all data taking. 
We would like to thank our technical staff: 
G.F.Fortugno for his dedicated work to ensure efficient operations of 
the KLOE computing facilities; 
M.Anelli for his continuous support of the gas system and the safety of the
detector; 
A.Balla, M.Gatta, G.Corradi and G.Papalino for the maintenance of the
electronics; 
M.Santoni, G.Paoluzzi and R.Rosellini for the general support the
detector; 
C.Piscitelli for his help during major maintenance periods.

This work was supported in part by DOE grant DE-FG-02-97ER41027; 
by EURODA$\Phi$NE contract FMRX-CT98-0169; 
by the German Federal Ministry of Education and Research (BMBF) contract 06-KA-957; 
by Graduiertenkolleg `H.E. Phys. and Part. Astrophys.' of Deutsche Forschungsgemeinschaft,
Contract No. GK 742; 
by INTAS, contracts 96-624, 99-37; 
and by the EU Integrated Infrastructure Initiative Hadron Physics
Project under contract number RII3-CT-2004-506078.

\end{document}